\UseRawInputEncoding
\documentclass[pra,superscriptaddress, notitlepage, reprint,twocolumn]{revtex4-1}
\usepackage[english]{babel}
\usepackage{amsmath,amsthm}
\usepackage{amsfonts}
\usepackage[pdfborder={0 0 0}, colorlinks=true, urlcolor=blue, linkcolor=blue, citecolor=blue]{hyperref}
\usepackage{color}
\usepackage{graphicx}
\usepackage{float}
\usepackage{subfigure}
\usepackage{lipsum}
\usepackage{epstopdf}
\usepackage{dcolumn}
\begin{document}

\title{Optimal temperature estimation in polariton Bose-Einstein Condensate}
\author{Dong  Xie}
\email{xiedong@mail.ustc.edu.cn}
\affiliation{College of Science, Guilin University of Aerospace Technology, Guilin, Guangxi 541004, People's Republic of China}

\begin{abstract}
Improving the measurement precision of temperature is very important and challenging, especially in the low temperature range. Based on the existence of invariant subspaces during the polariton thermalization, we propose a new way to enhance the measurement precision of the low temperature and obtain Landau bound to avoid that the measurement uncertainty of the temperature diverges as the temperature approaches zero. The measurement precision of the low temperature increases significantly with the number of polariton states. In order to resist the dissipation, the incoherent pumping is necessary for obtaining the information of the temperature encoded in the steady state. It should be noted that too strong incoherent pumping is wasteful  due to that the quantum Fisher information of the temperature becomes less and less dependent on the total number of the polaritons.
\end{abstract}
\maketitle

\textit{Introduction.-}Precise estimation of temperature is significant and crucial for the fundamental natural science and the changing quantum technology\cite{lab1a,lab2a,lab3a,lab1,lab2,lab3,lab4,lab5,lab6,lab7,lab8}. Since quantum devices generally work at low temperature\cite{lab9,lab10,lab11} and the development of the field of quantum thermodynamics\cite{lab12,lab13,lab14,lab15,lab16} also need the low temperature to reserve the quantum properties, precisely control and measurement of the low temperature is becoming an important subject in quantum metrology\cite{lab8} and quantum sensing.

Enhancing the estimation precision of temperature with quantum resources and investigating the fundamental limitations on temperature estimation have attracted a lot of attention\cite{lab17,lab18,lab19,lab20,lab21,lab22,lab23,lab24,lab25,lab26,lab27,lab28}. Temperature measurements can generally be divided into two categories: one is to measure the temperature encoded in the thermal equilibrium state of the probe system, the other is in the non-equilibrium state. Either way, the measurement of the low temperature has always been a challenging and arduous task. The uncertainty of the temperature diverges as the temperature approaches zero\cite{lab29}. Recently, some works attempted to alleviate the divergence. Correa \textit{et al.}\cite{lab30} showed that the thermometric precision at low temperature could be significantly enhanced by the strong probe-sampling coupling. Mukherjee \textit{et al.}\cite{lab31} utilized the periodic modulation to obtain the low-temperature thermometry with temperature-independent relative uncertainty. More importantly, Zhang \textit{et al.}\cite{lab32} obtained the Landau bound\cite{lab33} by using a continuous-variable system to detect the temperature of a non-Markovian reservoir.

In this work, we propose a new way to obtain the optimal low temperature estimation, and the Landau bound can also be achieved. The exciton-polariton Bose-Einstein Condensates (BECs) are used as the thermometer to measure the temperature of the phonons in the semiconductors or the temperature of intermolecular oscillations of the organic dyes. Different with the Bose polaron model\cite{lab9} in BEC, there are invariant subspaces due to the polariton thermalization, in which the final thermal state is not unique. For the infinite states of the polaritons, the Landau bound can also be achieved at the low temperature. In addition, we find that incoherent pumping can resist the dissipation. But too strong incoherent pumping is useless in enhancing  the estimation precision of the temperature due to that the temperature estimation precision will not increase with the increase of the total number of the polaritons for enough polaritons.

\textit{The physical model of polaritons.-}
When cavity photons strongly interact with an optical transition of active material, new eigenstates can be generated, i.e., lower and upper exciton-polariton branches\cite{lab34,lab35}. We only consider the lower branch in which the BEC occurs. Due to pair particle scattering, the exciton  polaritons of the lower polariton branch can be treated as harmonic oscillators by neglecting the nonlinearity\cite{lab36}.  The Hamiltonian of the polaritons can be described as
\begin{align}
H_{LP}=\sum_{j=0}^M\hbar\omega_j a^\dagger_j a_j.
\label{eq:1}
\end{align}
where the ground state (mode) is the state with $j=0$ and $a_j$ ($a_j^\dagger$) denotes the bosonic annihilation (creation) operator. The master equation for the the density matrix of the polaritons $\rho$, subject only to the polariton thermalization, can be expressed as
\begin{align}
\frac{d}{dt}\rho(t)=\mathcal{L}\rho=-i[H,\rho]+L_{\textmd{thermal}}(\rho),
\label{eq:2}
\end{align}
where the polariton thermalization is described by  the Lindblad superoperator
\begin{align}
&L_{\textmd{thermal}}(\rho)=\nonumber\\
&\sum_{j=0}^M\sum_{k=0}^M\Gamma_{jk}(a_ja_k^\dagger\rho a_k a_j^\dagger-\frac{1}{2}\rho a_k a_j^\dagger a_k a_j^\dagger-\frac{1}{2} a_k a_j^\dagger a_k a_j^\dagger\rho ),
\label{eq:3}
\end{align}
where $\Gamma_{jk}$ is the transition rate from the $j$th polariton state to the $k$th state. The thermalization rates obey the Kubo-Martin-Schwinger relation\cite{lab36a,lab37a} $\Gamma_{jk}=\exp[\frac{\hbar(\omega_j-\omega_k)}{\kappa_BT}]\Gamma_{kj}$, where $\kappa_B$ is the Boltzmann constant and $T$ is the temperature of intermolecular oscillations of the organic dyes or the temperature of the phonons in the semiconductors that we want to estimate. There are different underlying mechanisms of the polariton thermalization, which is dependent on the detail system. For example, the polariton thermalization comes from the nonlinear interaction with low frequency vibrations in organic polariton systems\cite{lab37,lab38}.

The operator of the total polariton number $\sum_{j=0}^Ma_j^\dagger a_j$ is the constant of motion during the thermalization process. The constant of motion implies that there are invariant subspaces $|n_0,n_1,...,n_M\rangle\langle n_0,n_1,...,n_M|$ with the total number of polaritons equal to $\sum_{j=0}^Mn_j=N$. When there are invariant subspaces, the stationary solution is not unique\cite{lab39}. The Gibbs distribution over the states of a given invariant subspace is also a stationary solution, which is given by\cite{lab36}
\begin{align}
\rho_s=\sum_{N=0}^\infty P_N(0)\frac{1}{Z_N}\sum_{n_0+...+n_M=N}\nu_0^{n_0}...\nu_M^{n_M}\nonumber \\
\times|n_0,n_1,...,n_M\rangle\langle n_0,n_1,...,n_M|,
\label{eq:4}
\end{align}
where $P_N(0)$ denotes the probability that there are $N$ polaritons in total in the low polariton branch at the initial time, which is given by $P_N(0)=\sum_{n_0+...+n_M=N}\textmd{Tr}[\rho(0)|n_0,n_1,...,n_M\rangle\langle n_0,n_1,...,n_M|]$ with $\sum_{j=0}^Mn_j=N$, $\nu_j=\exp[\frac{\hbar (\omega_0-\omega_j)}{k_BT}]$, and $Z_N$ is the partition function discribed by
\begin{align}
Z_N=\sum_{n_0+...+n_M=N}\nu_0^{n_0}...\nu_M^{n_M}.
\label{eq:5}
\end{align}

The estimation uncertainty of the unbiased estimator $\delta^2T$ is bounded by the quantum
Cram\'{e}r-Rao lower bound as $\delta^2T\geq1/F_T$\cite{lab40,lab41}, where $F_T$ is the quantum Fisher information (QFI)\cite{lab42} of temperature $T$ in the steady state $\rho_s$, which is given by
\begin{align}
F_T=\sum_{N=0}^\infty \sum_{n_0+...+n_M=N}\frac{(\partial_T P_{N,n_0,..,n_M})^2}{ P_{N,n_0,..,n_M}},
\label{eq:6}
\end{align}
where $P_{N,n_0,..,n_M}=\frac{P_N(0)}{Z_N}\nu_0^{n_0}...\nu_M^{n_M}$ denotes the probability of  projection into the state $|n_0,n_1,...,n_M\rangle$ with $\sum_{j=0}^Mn_j=N$, and the shorthand notation $\partial_T=\frac{\partial}{\partial T} $. Hereafter, we set $\hbar=\kappa_B=1$
for convenience.

Without loss of generality, we first consider that the system is consisted of $M+1$ states equidistant in frequency, i.e., $\omega_j=\omega_0+\omega\times j/M$ with $\omega=\omega_M-\omega_0$.

\textit{In the case of two polariton states.-}For $M=1$, we can obtain the probability of the state $|n_0,n_1\rangle$ with $n_0+n_1=N$
\begin{align}
P_{N,n_0,n_1}=\langle n_0,n_1|\rho_s(M=1)|n_0,n_1\rangle=\frac{\lambda^{-n_1}(\lambda-1)}{\lambda-\lambda^{-N}},
\end{align}
where the factor $\lambda$ is defined as $\lambda=\exp(\omega/T)$, and only one invariant subspace is considered, i.e., $P_N(0)=1$.
By utilizing the formula in Eq.~(\ref{eq:6}), the QFI with $M=1$ is analytically expressed as
\begin{align}
&F_T=\frac{\omega^2}{T^4}[\frac{1}{\lambda-1}+\frac{1}{(\lambda-1)^2}-\frac{(1+N)^2}{\lambda^{1+N}-1}-\frac{(1+N)^2}{(\lambda^{1+N}-1)^2}].
\label{eq:8}
\end{align}

For $N\gg1$, we can obtain the simplified form $F_T\simeq\frac{\lambda \omega^2}{(\lambda-1)^2T^4}$. This result is equivalent to measuring the temperature $T$ in the thermal equilibrium state of harmonic oscillator with the Hamiltonian $H_e=\omega a^\dagger a$\cite{lab25}. This does not reflect the advantage of having invariant subspaces. This is mainly due to the low number of states. Next, we investigate the QFI of the temperature with a large number of states, i.e., $M>1$.

\begin{figure}
  \includegraphics[scale=0.6]{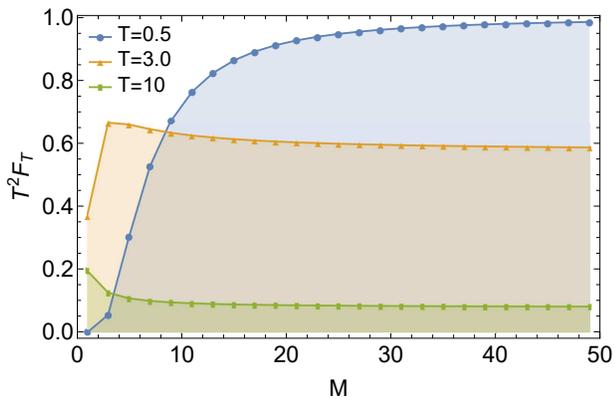}
  \caption{\label{fig.1}The change relation diagram of the ratio of the QFI corresponding to finite $M$ value and the maximal QFI $1/T^2$ corresponding to infinite $M$,  i.e., $T^2F_T$, versus $M$ for three different temperatures in arbitrary units: $T=0.5,\ 3.0,\ 10$. Here the dimensionless parameters are given by $\omega=10$ and $N=1$.}
 \end{figure}

\textit{In the case of single polarization: $N=1$.-} We then consider another simple case, which can be analytically calculated  for $N=1$.
In this case, the probability of projection into the state $|0,0,...,n_j=1,...0\rangle$ is $P_{N=1,n_j=1}=\frac{\lambda^{-j/M}}{\sum_{j=0}^M\lambda^{-j/M}}=\frac{\lambda^{1-j/M}(\lambda^{1/M}-1)}{\lambda^(1+1/M)-1}$. The corresponding QFI is given by
\begin{align}
&F_T=2\omega^2\lambda^{1+2/M}\times\nonumber\\
&\frac{M(2+M)-(1+M)^2\cosh(\frac{\omega}{MT})+\cosh[(1+M)\frac{\omega}{MT}]}{(\lambda^{1/M}-1)^2[\lambda^{(1+M)/M}-1]^2M^2T^4}.
\label{eq:9}
\end{align}
For the infinite $M$, the QFI of the temperature $T$ tends to be
\begin{align}
F_T|_{M\rightarrow\infty}=\frac{1}{T^2}-\frac{\lambda \omega^2}{(\lambda-1)^2T^4}.
\end{align}
For $\omega\gg T$, we can obtain $F_T|_{M\rightarrow\infty}\simeq 1/T^2$, which is the maximal QFI of the temperature. According to the quantum Cram\'{e}r-Rao lower bound, we obtain the Landau bound\cite{lab33}, i.e., $\delta T\geq T$. It means that we can obtain the infinite QFI for the temperature $T\rightarrow0$, leading to that the optimal estimation precision $\delta T\approx0$ is achieved. The result shows that we can perform very accurate low temperature measurement in the case of the invariant subspace with infinite states. As shown in Fig.~\ref{fig.1}, the QFI increases with the value of $M$ for $\omega\gg T$. On the contrary, the QFI decreases with the value of $M$ for high temperature. For $M$ is larger than a certain characteristic value $M_c$, the relationship between QFI and $M$ becomes increasingly independent. From Fig.~\ref{fig.1}, we can see that the characteristic value $M_c$ decreases as the temperature $T$ increases. In the low temperature region, the increasing number of polariton states $M$ promotes the estimation precision of temperature more obviously.

\begin{figure}
  \includegraphics[scale=0.85]{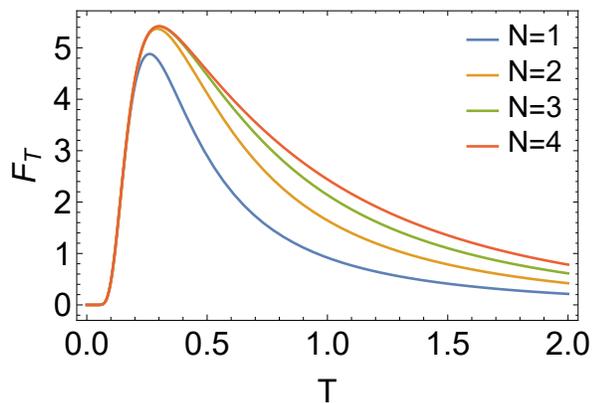}
  \caption{\label{fig.2}The QFI of the temperature changes with the temperature $T$ for the four numbers of polaritons: $N=1,\ 2,\ 3,\ 4$. Here the dimensionless parameters are given by $\omega=10$ and $M=10$.}
 \end{figure}

\begin{figure}
  \includegraphics[scale=0.85]{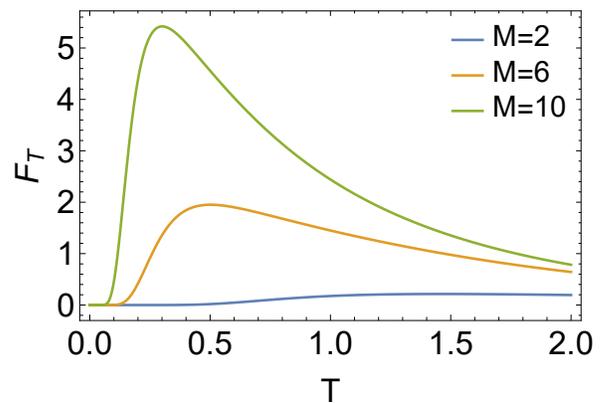}
  \caption{\label{fig.3}The QFI of the temperature changes with the temperature $T$ for the three numbers of polariton states: $M=2,\ 4,\ 6$. Here the dimensionless parameters are given by $\omega=10$, and $N=4$.}
 \end{figure}

For the general case, i.e., $N>1$ and $M>1$, there are no analytical solutions. It can be calculated numerically to investigate the effects of the total number of polaritons $N$ and the number of polariton states $M+1$ on the QFI of the temperature $T$.
As shown in Fig.~\ref{fig.2}, with  the  increase of the polaritons $N$, the QFI of the temperature $T$ increases when the temperature $T$ above a certain value. However, the enhancement effect becomes less and less obvious as $N$ increases. This result is consistent with the previous analytical result of $M=1$, which is independent of the total number of polaritons $N$ for large number $N\gg1$. By contrast, increasing $M$ can significantly improve the QFI with $N=4$, especially in the low-temperature areas, as shown in Fig.~\ref{fig.3}. This conclusion also supports the previous analytical results.

As a summary, one of our main results is that increasing the number of polariton states can greatly enhance the measurement precision of the low temperature. However, the total number of polaritons $N$ plays a smaller and smaller role in enhancing the measurement precision of the temperature as $N$ increases.

\textit{The polariton dissipation and incoherent pumping.-} We consider that there is a polariton dissipation in the polariton system, which is generally unavoidable in real quantum systems. We assume that the coupling between the polariton system and the environment is weak, leading to that the Born-Markov approximation\cite{lab42a,lab43a} can be utilized. Therefore, the polariton dissipation can be described by the Lindblad superoperator
\begin{align}
L_{\textmd{diss}}(\rho)=\sum_{j=0}^M\gamma_{j}(a_j\rho  a_j^\dagger-\frac{1}{2}\rho  a_j^\dagger a_j-\frac{1}{2}  a_j^\dagger a_j\rho ),
\end{align}
where $\gamma_{j}$ is the dissipation rate of the $j$th polariton state. Without extra pumping, the total number of the polaritons in the steady state will be 0 due to the dissipation, i.e., the steady system is in the vacuum state $|0,0,...,0\rangle$. The information of temperature can not be obtained by the steady state due to that the dissipation erases all information. In order to get the information of the temperature, it is necessary to take the measurements in advance before the system reaches the steady state. In this work, we assume that
the thermalization process is much faster than the dissipation process, i.e., $\Gamma_{0j}(1+\langle a_0^\dagger a_0\rangle)\gg\ \gamma_j$. The density matrix in Eq.~(\ref{eq:4}) can be obtained approximately at time $\gamma_j^{-1}\gg t\gg\Gamma_{0j}^{-1}(1+\langle a_0^\dagger a_0\rangle)^{-1}$. In thus case, the optimal interrogation time should be much smaller than the characteristic time of the dissipation process and larger than that of the thermalization process.

In order not to control the interrogation time, extra pumping is required to obtain a non-vacuum steady state. We consider that there is a incoherent pumping. The energy transfers from the upper branch and uncoupled excitons towards the lower branch can be treated as an effective incoherent pumping,  which can be described by the Lindblad master equation
\begin{align}
&L_{\textmd{pump}}(\rho)=\sum_{j=0}^M \kappa_{j}(a_j\rho  a_j^\dagger-\frac{1}{2}\rho  a_j^\dagger a_j-\frac{1}{2}  a_j^\dagger a_j\rho )+\nonumber\\
&\sum_{j=0}^M\kappa_{j}(a_j^\dagger\rho  a_j-\frac{1}{2}\rho  a_j a_j^\dagger-\frac{1}{2}  a_j a_j^\dagger\rho ),
\end{align}
where $\kappa_j$ is the pumping rate of the $j$th polariton state.
Including the dissipation and the incoherent pumping, the density matrix of the polaritons $\rho$ is dominated by the master equation
\begin{align}
\frac{d}{dt}\rho=-i[H,\rho]+L_{\textmd{thermal}}(\rho)+L_{\textmd{diss}}(\rho)+L_{\textmd{pump}}(\rho).
\label{eq:2}
\end{align}
In general, the above equation is difficult to be solved numerically and analytically. It can be approximately solved by assuming that the thermalization is much faster than the dissipation and the incoherent pumping, i.e., $\Gamma_{0j}(1+\langle a_0^\dagger a_0\rangle)\gg
\gamma_j,\kappa_j$\cite{lab43,lab44}. The general expression for the density matrix in the steady state $\rho_s$ is also described by
 \begin{align}
\rho_s=\sum_{N=0}^\infty P_N\frac{1}{Z_N}\sum_{n_0+...+n_M=N}\nu_0^{n_0}...\nu_M^{n_M}\nonumber \\
\times|n_0,n_1,...,n_M\rangle\langle n_0,n_1,...,n_M|,
\label{eq:2}
\end{align}
where $P_N$ denotes the probability that there are $N$ polaritons in total in the steady state.  In this case, $P_N$ is independent of the initial value $P_N(0)$. Without loss of generality, we consider $\kappa_j=\kappa$ and $\gamma_j=\gamma$. For only two polariton states ($M=1$), we can obtain the general form of the probabilities (see Appendix A for details)
\begin{align}
P_N=(\frac{\kappa}{\kappa+\gamma})^N(N+1)P_0, \\
P_0=\frac{1}{1+\frac{\kappa(\kappa+2\gamma)}{\gamma^2}}.
\end{align}

\begin{figure}
  \includegraphics[scale=0.8]{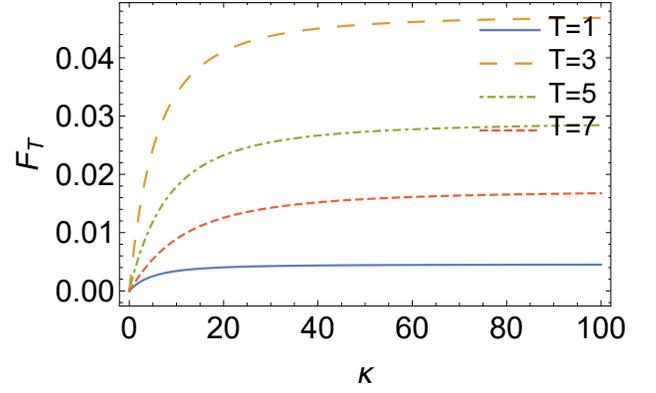}
  \caption{\label{fig.4}Diagram of the QFI $F_T$ as a function of pumping rate $\kappa$ with four different temperature: $T=1,3,5,7$. Here, the dimensionless parameters are given by $\omega=10$, and $\gamma=10$.}
 \end{figure}

In the steady state, the probability of the state $|n_0,n_1\rangle$ with $n_0+n_1=N$ is given by
 \begin{align}
P_{N,n_0,n_1}=P_N\lambda^{-n_1}(\lambda-1)/(\lambda-\lambda^{-N}),
\label{eq:17}
\end{align}
where $n_1$ ranges from $0$ to $N$.
The corresponding QFI can be calculated by combing Eq.~(\ref{eq:6}) and Eq.~(\ref{eq:17}), and the results are directly shown in Fig.~\ref{fig.4}. The QFI can increase with the pumping rate $\kappa$ for different temperature, but not all the time. When the pumping rate $\kappa$ is larger than a certain value, the QFI is independent of the pumping rate. It is because that the number of polaritons $N$ is significantly increased by the strong enough pumping, i.e., $\sum_{N\gg1}P_{N}\sim1$. As shown in Eq.~(\ref{eq:8}), the QFI is independent of the number of polaritons $N$ in the case of $N\gg1$. Therefore, too strong pumping is wasteful in improving the measurement precision of the temperature $T$. Only the appropriate strength of the incoherent pumping is required to resist the effects of dissipation.

\textit{M degenerate states.-}
In order to deal with the case of $N\gg1$ and  $M\gg1$, we consider that there are $M$ degenerate states (modes), i.e., $\omega_j=\omega_0+\omega$ for $j\geq1$.
The general form of the partition function $Z^d_N$ in the case of $M$ degenerate states are obtained (see Appendix B for details)
 \begin{align}
 Z^d_{N}=\frac{\lambda^M}{(\lambda-1)^M} - \frac{
 (M+N)!_2F_1[1, 1+M+N,2+N;\frac{1}{\lambda}]}{(M-1)! (N+1)!\lambda^{(1 + N)}},
\label{eq:18}
\end{align}
where $_2F_1(a,b,c;d)=\sum_{k=0}^\infty\frac{(a)_k(b)_k d^k}{(c)_k k!}$ denotes the generalized Hypergeometric function with $(r)_k=\Gamma(r+k)/\Gamma(r)$.
The general form of the probability of the state $|n_0,...,n_M\rangle$ with $n_0+...+n_M=N$ is given by
\begin{align}
P^d_{N,n_0,..,n_M}=\frac{\lambda^{n_0-N}P^d_N}{Z_N},
\end{align}
where $P^d_N$ denotes the probability that there are $N$ polaritons in total in the steady state: $P^d_N=P_N(0)$ in the absence of  the dissipation and the incoherent pumping; $P^d_N=P_N$ with the dissipation and the incoherent pumping.
We note that there are $(N-n_0 + M - 1)!/[(N-n_0)! (M - 1)!]$ different states $|n_0,...,n_M\rangle$ with the same probability $P^d_{N,n_0,..,n_M}$ due to the $M$ degenerate modes. This simplifies the calculation somewhat. We numerically calculate the corresponding QFI $F_T^d$ with the large $N$ and $M$.
\begin{figure}
  \includegraphics[scale=0.75]{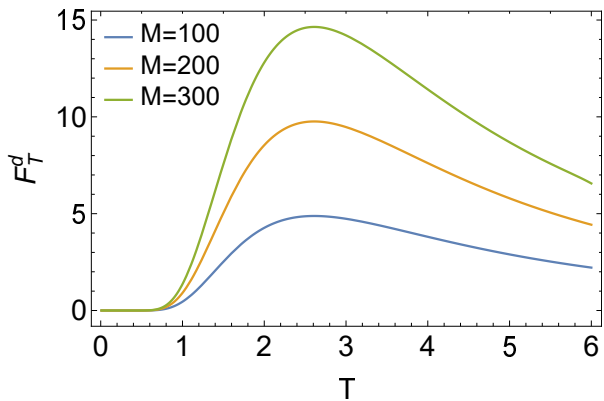}
  \caption{\label{fig.5}Diagram of the QFI $F^d_T$ as the function of the temperature $T$ with three different values of $M$: $M=100,200,300$.  The number of the polaritons is fixed, i.e., $P_N^d=1$. Here, the dimensionless parameters are given by $\omega=10$, and $N=100$.}
 \end{figure}

As shown in Fig.~\ref{fig.5}, the QFI increases with the number of states $M$ with the fixed number of the polaritons $N=100$, especially in the low temperature. It means that the number of polariton states $M$ is an important resource for enhancing the
estimation precision of the temperature, which is consistent with the previous results by using the small $N$ and $M$ in the case of states equidistant in frequency.

\begin{figure}
  \includegraphics[scale=0.85]{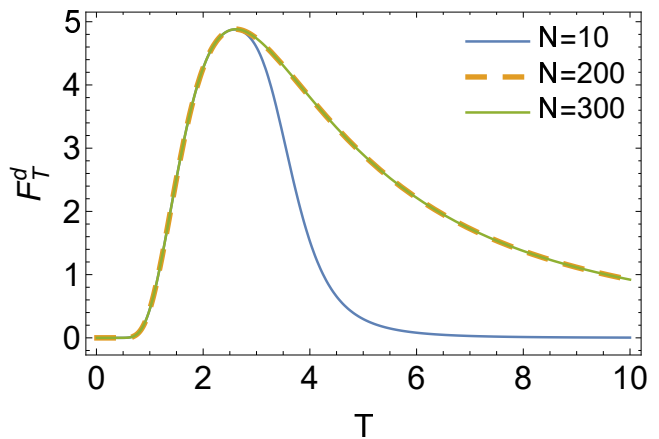}
  \caption{\label{fig.6}Diagram of the QFI $F_T$ as the function of the temperature $T$ with three different numbers of the polaritons: $N=10,200,300$. Here, the dimensionless parameters are given by $\omega=10$, $P_N^d=1$, and $M=100$.}
 \end{figure}

 As shown in Fig.~\ref{fig.6}, in the case of large $M$,
the line $N=100$ is coincident with the line $N=200$. It implies that the QFI of the temperature is independent of the number of the polaritons for large $N$ in the case of large $M$.
The value of $M$ can not change the relation between the QFI and the number of the polaritons $N$.
\textit{Discussion and Conclusion.-}
We have proposed a new mechanism to improve the measurement precision of the low temperature and obtained Landau bound $\delta T\simeq T$.
It is based on the existence of invariant subspaces during the polariton thermalization. The measurement precision of the low temperature increases significantly with the number of polariton states. It is interesting to apply our mechanism to different systems beyond polariton BECs by looking for invariant subspaces with many modes during the thermalization.

When the polaritons suffer from the dissipation, the incoherent pumping can be used to enhancing the estimation precision of temperature. Due to that the QFI becomes less and less dependent on the total number of the polaritons, too strong incoherent pumping is a waste of energy. Whether the periodic coherent modulation and nonlinearity can be used to enhancing the estimation precision of low tempearature deserves further study.

 \textit{Acknowledgements.-}This research was supported by the National Natural Science Foundation of China under Grant No. 62001134, Guangxi Natural Science Foundation under Grant No. 2020GXNSFAA159047 and National Key R\&D Program of China under Grant No. 2018YFB1601402-2.

\section*{appendix A}
We consider that there are polariton dissipation and incoherent pumping of the lower polariton states, which are described by the Lindblad superoperators

\begin{align}
&L_{\textmd{diss}}(\rho)=\sum_{j=0}^M\gamma_{j}(a_j\rho  a_j^\dagger-\frac{1}{2}\rho  a_j^\dagger a_j-\frac{1}{2}  a_j^\dagger a_j\rho )\tag{S1}\\
&L_{\textmd{pump}}(\rho)=\sum_{j=0}^M \kappa_{j}(a_j\rho  a_j^\dagger-\frac{1}{2}\rho  a_j^\dagger a_j-\frac{1}{2}  a_j^\dagger a_j\rho )+\nonumber\\
&\sum_{j=0}^M\kappa_{j}(a_j^\dagger\rho  a_j-\frac{1}{2}\rho  a_j a_j^\dagger-\frac{1}{2}  a_j a_j^\dagger\rho ),\tag{S2}
\end{align}
where $\gamma_j$ ($\kappa_j$) is the dissipation (the incoherent pumping) rate of the $j$th state.
The general solution is difficult and we also consider that the thermalization process is very rapid, i.e., $\Gamma_{0j}(1+
\langle a_0^\dagger a_0\rangle)\gg\kappa_j,\ \gamma_j$, which has been proved to be valid in ref.\cite{lab36}.
 In the first stage, i.e., $\kappa_j^{-1},\ \gamma_j^{-1}\gg t\gg \Gamma_{0j}^{-1}(1+ \langle a_0^\dagger a_0\rangle)^{-1}$, the density matrix $\rho(t)$ obeys the approximate differential equation by ignoring the effect of the dissipation and the incoherent pumping,
 \begin{align}
\frac{d}{dt}\rho\approx-i[H,\rho]+L_{\textmd{thermal}}(\rho),
\tag{S3}
\end{align}
  which leads to the general expression for the density matrix $\rho$\cite{lab36}
  \begin{align}
\rho=\sum_{N=0}^\infty P_N(t)\frac{1}{Z_N}\sum_{n_0+...+n_M=N}\nu_0^{n_0}...\nu_M^{n_M}\nonumber \\
\times|n_0,n_1,...,n_M\rangle\langle n_0,n_1,...,n_M|,
\tag{S4}
\label{eq:S4}
\end{align}
where $P_N(t)\approx P_N(0)$ for $\kappa_j^{-1},\ \gamma_j^{-1}\gg t\gg \Gamma_{0j}^{-1}(1+ \langle a_0^\dagger a_0\rangle)^{-1}$.

In the second stage, one can substitute the above equation into the whole evolution equation of the density matrix $\rho$, which is described as
\begin{align}
\frac{d}{dt}\rho=-i[H,\rho]+L_{\textmd{thermal}}(\rho)+L_{\textmd{diss}}(\rho)+L_{\textmd{pump}}(\rho).
\tag{S5}
\label{eq:S5}
\end{align}

Substituting Eq.~(\ref{eq:S4}) into Eq.~(\ref{eq:S5}) and using the steady conditions $\frac{d}{dt}P_N(t\rightarrow\infty)=0$,
 the steady-state solutions of the probabilities $P_N=P_N(t\rightarrow\infty)$ are achieved\cite{lab36}
\begin{align}
P_{N+1}=\frac{(d_{N-1}+ \beta_{N})Z_{N+1}}{d_N Z_{N}}P_{N}-\frac{\beta_{N-1} Z_{N+1}}{d_N Z_{N-1}}P_{N-1},\tag{S6}\\
P_1=\frac{\beta_0 Z_1}{d_0 Z_0}P_0,
\tag{S7}
\end{align}
where $d_N=\sum_{n=0}^N\sum_{j=0}^M(\gamma_j+\kappa_j)\nu_j^{n+1}Z_{N-n}$ and $\beta_N=\sum_{n=0}^N\sum_{j=0}^M\kappa_j\nu_j^{n}Z_{N-n}$.
Utilizing the above equations, we further obtain that
\begin{align}
P_2=\frac{\beta_1 Z_2}{d_1 Z_1}P_1,\tag{S8}\\
P_3=\frac{\beta_2 Z_3}{d_2 Z_2}P_2,\tag{S9}\\
P_4=\frac{\beta_3 Z_4}{d_3 Z_3}P_3,\tag{S10}\\
P_5=\frac{\beta_4 Z_5}{d_4 Z_4}P_4,\tag{S11}\\
...\nonumber
\end{align}
By induction, we can obtain the general form of the probability $P_N$ with $N\geq1$
\begin{align}
P_N=\frac{\beta_{N-1} Z_{N}}{d_{N-1} Z_{N-1}}P_{N-1}\tag{S12}\\
=\frac{\prod_{n=0}^{N-1}\beta_{n} Z_{N}}{\prod_{n=0}^{N-1}d_{n}}P_0.\tag{S13}
\label{eq:S13}
\end{align}

Without loss of generality, we consider $\kappa_j=\kappa$ and $\gamma_j=\gamma$. In the case of two states ($M=1$), we can obtain the analytical results about the distribution
\begin{align}
d_N=\frac{\lambda^{-(1+N)}(\lambda^{2+N}-1)(1+N)(\kappa+\gamma)}{(\lambda-1)},\tag{S14}\\
\beta_N=\frac{\lambda^{-N}(\lambda^{1+N}-1)(2+N)\kappa}{(\lambda-1)},\tag{S15}\\
Z_N=\sum_{n=0}^N\exp[-n \omega/T]=\frac{\lambda-\lambda^{-N}}{\lambda-1}.\tag{S16}
\end{align}
Substituting the above equations into Eq.~(\ref{eq:S13}), we can derive a simplified form of the probability $P_N$, which is described as
\begin{align}
P_N=(\frac{\kappa}{\kappa+\gamma})^N(N+1)P_0.
\tag{S17}
\label{eq:S17}
\end{align}

Due to the normalization condition $\sum_{N=0}^\infty P_N=1$, we can obtain that
\begin{align}
P_N=(\frac{\kappa}{\kappa+\gamma})^N(N+1)P_0, \tag{S18}\\
P_0=\frac{1}{1+\frac{\kappa(\kappa+2\gamma)}{\gamma^2}}.\tag{S19}
\end{align}

\section*{appendix B}
In the case of $M$ degenerate states, $\omega_j=\omega_0+\omega$ for $j\geq1$,  the corresponding
density matrix in the steady state $\rho^d_s=\rho^d(t\rightarrow\infty)$ is also written as
\begin{align}
\rho^d_s=\sum_{N=0}^\infty P^d_N\frac{1}{Z^d_N}\sum_{n_0+...+n_M=N}\nu_0^{n_0}...\nu_M^{n_M}\nonumber \\
\times|n_0,n_1,...,n_M\rangle\langle n_0,n_1,...,n_M|,\tag{S20}
\end{align}
The partition function can be derived
\begin{align}
Z^d_N=\sum_{n_0+...+n_M=N}\lambda^{-(n_1+...+n_M)}=\sum_{n_0=0}^N\mu(n_0)\lambda^{n_0-N}\tag{S21}\\
=\frac{\lambda^M}{(\lambda-1)^M}-\frac{
 (M+N)!_2F_1[1, 1+M+N,2+N;\frac{1}{\lambda}]}{(M-1)! (N+1)!\lambda^{(1 + N)}},\tag{S22}
\end{align}
where the degeneration coefficient $\mu(n_0)=(N-n_0+M-1)!/[N!(M-1)!]$ comes from the $M$ degenerate modes, and  $_2F_1(a,b,c;d)=\sum_{k=0}^\infty\frac{(a)_k(b)_k d^k}{(c)_k k!}$ denotes the generalized Hypergeometric function with $(r)_k=\Gamma(r+k)/\Gamma(r)$.
The general form of the probability of the state $|n_0,n_1,...,n_M\rangle$ with the total number $n_0+n_1...+n_M=N$ is given by
\begin{align}
P^d_{N,n_0,..,n_M}=\frac{\lambda^{n_0-N}P^d_N}{Z^d_N},\tag{S23}
\end{align}
where $P^d_N$ denotes the probability that there are $N$ polaritons in total in the steady state: $P^d_N=P_N(0)$ in the absence of the dissipation and the incoherent pumping; $P^d_N=P_N$ with the dissipation and the incoherent pumping. In the case of $M$ degenerate states, the general formula of QFI $F^d_T$ can be described as
\begin{align}
F^d_T=\sum_{N=0}^\infty \sum_{n_0+...+n_M=N}\frac{(\partial_T P^d_{N,n_0,..,n_M})^2}{ P^d_{N,n_0,..,n_M}}\tag{S24}\\
=\sum_{N=0}^\infty \sum_{n_0=0}^N\mu(n_0)\times\frac{(\partial_T \frac{\lambda^{n_0-N}P^d_N}{Z^d_N})^2}{ \frac{\lambda^{n_0-N}P^d_N}{Z^d_N}}\tag{S25}\\
=\sum_{N=0}^\infty \sum_{n_0=0}^N\mu(N-n_0)\times\frac{(\partial_T \frac{\lambda^{-n_0}P^d_N}{Z^d_N})^2}{ \frac{\lambda^{-n_0}P^d_N}{Z^d_N}}.
\tag{S26}
\end{align}

For the fixed number of the polaritons, i.e., $P^d_N=1$, the QFI can be calculated by
\begin{align}
F^d_T= \sum_{n_0=0}^N\mu(N-n_0)\times\frac{(\partial_T \frac{\lambda^{-n_0}}{Z^d_N})^2}{ \frac{\lambda^{-n_0}}{Z^d_N}}.
\tag{S27}
\end{align}


\begin{thebibliography}{9}

\vspace{3mm}
\bibitem{lab1a}Y. Gao, and Y. Bando, Nanotechnology: carbon nanothermometer containing gallium, Nature 415, 599 (2002).
\bibitem{lab2a}D. M. Weld, P. Medley, H. Miyake, D. Hucul,  D. E. Pritchard, and W. Ketterle, Spin gradient thermometry for ultracold atoms in optical lattices, Phys. Rev. Lett. 103, 245301 (2009).
\bibitem{lab3a}P. Neumann, I. Jakobi, F. Dolde, C. Burk, R. Reuter, G. Waldherr, J. Honert, T. Wolf, A. Brunner,and J. H. Shim, High-precision nanoscale temperature sensing using single defects in diamond. Nano Lett. 13, 2738 (2013).
\bibitem{lab1}F. Giazotto, T. T. Heikkil\"{a}, A. Luukanen, A. M. Savin, and
J. P. Pekola, Opportunities for mesoscopics in thermometry and refrigeration: Physics and applications, Rev. Mod.
Phys. 78, 217 (2006).
\bibitem{lab2}N. Navon, S. Nascimb\`{e}ne, F. Chevy, and C. Salomon, The Equation of State of a Low-Temperature Fermi Gas with Tunable Interactions, Science 328, 729 (2010).
\bibitem{lab3}M. J. H. Ku, A. T. Sommer, L. W. Cheuk, and M. W. Zwierlein, Revealing the Superfluid Lambda Transition in the Universal Thermodynamics of a Unitary Fermi Gas, Science 335, 563 (2012).
\bibitem{lab4}L. D. Carlos, F. Palacio, eds., Thermometry at the
Nanoscale (The Royal Society of Chemistry, Cambridge,
2016).
\bibitem{lab5}W. Hofstetter and T. Qin, Quantum simulation of strongly correlated condensed matter systems, J. Phys. B 51, 082001 (2018).
\bibitem{lab6}L. Tarruell and L. Sanchez-Palencia, Quantum simulation of the Hubbard model with ultracold fermions in optical lattices, C. R. Phys. 19, 365 (2018).
\bibitem{lab7} A. De Pasquale, and T. M. Stace, in Thermodynamics in the
Quantum Regime: Fundamental Aspects and New Directions, edited by F. Binder, L. A. Correa, C. Gogolin, J.
Anders, and G. Adesso (Springer International Publishing,
Cham, 2018), p. 503.
\bibitem{lab8}M. Mehboudi, A. Sanpera, and L. A. Correa, Thermometry
in the quantum regime: Recent theoretical progress, J. Phys.
A: Math. Theor. 52, 303001 (2019).
\bibitem{lab9}M. Mehboudi, A. Lampo, C. Charalambous, L. A. Correa,
M. A. Garc¨ªa-March, and M. Lewenstein, Using Polarons
for sub-NK Quantum Nondemolition Thermometry in a
Bose-Einstein Condensate, Phys. Rev. Lett. 122, 030403
(2019).
\bibitem{lab10}Q. Bouton, J. Nettersheim, D. Adam, F. Schmidt, D. Mayer,
T. Lausch, E. Tiemann, and A. Widera, Single-Atom Quantum Probes for Ultracold Gases Boosted by Nonequilibrium
Spin Dynamics, Phys. Rev. X 10, 011018 (2020).
\bibitem{lab11}M. T. Mitchison, T. Fogarty, G. Guarnieri, S. Campbell, T.
Busch, and J. Goold, In Situ Thermometry of a Cold Fermi
gas via Dephasing Impurities, Phys. Rev. Lett. 125, 080402
(2020).
\bibitem{lab12}M. Campisi, P. H\"{a}nggi, and P. Talkner, Colloquium: Quantum fluctuation relations: Foundations and applications,
Rev. Mod. Phys. 83, 771 (2011).
\bibitem{lab13}F. Brand\~{a}o, M. Horodecki, N. Ng, J. Oppenheim, and S.
Wehner, The second laws of quantum thermodynamics,
Proc. Natl. Acad. Sci. USA 112, 3275 (2015).
\bibitem{lab14}S. Vinjanampathy and J. Anders, Quantum thermodynamics, Contemp. Phys. 57, 545 (2016).
\bibitem{lab15}S. Deffner and S. Campbell, Quantum Thermodynamics
(Morgan \& Claypool Publishers, 2019).
\bibitem{lab16}U. Marzolino and D. Braun, Precision measurements of
temperature and chemical potential of quantum gases, Phys.
Rev. A 88, 063609 (2013).
\bibitem{lab17}D. Xie, C. Xu, and A. Wang, Optimal quantum thermometry by dephasing, Quantum Inf Process 16, 155 (2017).
\bibitem{lab18}J. Yang, C. Elouard, J. Splettstoesser, B. Sothmann, R. S\'{a}nchez, and A. N. Jordan, Thermal transistor and thermometer based on Coulomb-coupled conductors, Phys. Rev. B 100, 045418 (2019).
\bibitem{lab19}P. P. Hofer, J. B. Brask, M. Perarnau-Llobet, and N. Brunner, Quantum Thermal Machine as a Thermometer,
Phys. Rev. Lett. 119, 090603 (2017).
\bibitem{lab20}L. Spietz, K. W. Lehnert, I. Siddiqi, and R. J. Schoelkopf, Primary Electronic Thermometry Using the Shot Noise of a Tunnel Junction, Science 300, 1929 (2003).
\bibitem{lab21}L. Spietz, R. J. Schoelkopf, and P. Pari, Shot noise thermometry down to 10mK, Appl. Phys. Lett. 89,
183123 (2006).
\bibitem{lab22}S. Gasparinetti, F. Deon, G. Biasiol, L. Sorba, F. Beltram, and
F. Giazotto, Probing the local temperature of a two-dimensional electron gas microdomain with a quantum dot: Measurement of electron-phonon interaction, Phys. Rev. B 83, 201306(R) (2011).
\bibitem{lab23}M. Brunelli, S. Olivares, and M. G. A. Paris, Qubit thermometry for micromechanical resonators, Phys. Rev. A 84,
032105 (2011).
\bibitem{lab24}S. Jevtic, D. Newman, T. Rudolph, and T. M. Stace, Single qubit thermometry, Phys. Rev. A 91, 012331 (2015).
\bibitem{lab25}L. A. Correa, M. Mehboudi, G. Adesso, and A. Sanpera,
Individual Quantum Probes for Optimal Thermometry,
Phys. Rev. Lett. 114, 220405 (2015).
\bibitem{lab26}S. Campbell, M. Mehboudi, G. D. Chiara, and M.
Paternostro, Global and local thermometry schemes in
coupled quantum systems, New J. Phys. 19, 103003
(2017).
\bibitem{lab27}A. H. Kiilerich, A. De Pasquale, and V. Giovannetti,
Dynamical approach to ancilla-assisted quantum thermometry, Phys. Rev. A 98, 042124 (2018).
\bibitem{lab28}D. Xie, F. Sun, and C. Xu, Quantum thermometry based on a cavity-QED setup, Phys. Rev. A 101, 063844 (2020).
\bibitem{lab29}G. De Palma, A. De Pasquale, and V. Giovannetti, Universal locality of quantum thermal susceptibility, Phys. Rev. A
95, 052115 (2017).
\bibitem{lab30}Luis A. Correa, Mart¨ª Perarnau-Llobet, Karen V. Hovhannisyan, Senaida Hern¨¢ndez-Santana,
Mohammad Mehboudi, and Anna Sanpera, Enhancement of low-temperature thermometry by strong coupling, Phys. Rev. A 96, 062103 (2017).
\bibitem{lab31}V. Mukherjee, A. Zwick, A. Ghosh, X. Chen, and G. Kurizki, Enhanced precision bound of low-temperature quantum thermometry via dynamical control, Commun. Phys.
2, 162 (2019).
\bibitem{lab32}Ning Zhang, Chong Chen, Si-Yuan Bai, Wei Wu, and Jun-Hong An, Non-Markovian Quantum Thermometry, Phys. Rev. Applied 17, 034073 (2022).
\bibitem{lab33}M. G. A. Paris, Achieving the landau bound to precision
of quantum thermometry in systems with vanishing gap, J.
Phys. A: Math. Theor. 49, 03LT02 (2015).
\bibitem{lab34}T. Byrnes, N. Y. Kim, and Y. Yamamoto, Exciton-polariton condensates, Nat. Phys. 10, 803
(2014).
\bibitem{lab35}A. V. Zasedatelev, A. V. Baranikov, D. Urbonas, F. Scafirimuto, U. Scherf, T. St\"{o}ferle, R. F. Mahrt, and P. G.
Lagoudakis, A room-temperature organic polariton transistor, Nat. Photonics 13, 378 (2019).
\bibitem{lab36}Vladislav Yu. Shishkov and Evgeny S. Andrianov, Exact Analytical Solution for the Density Matrix of a Nonequilibrium
Polariton Bose-Einstein Condensate, Phys. Rev. Lett. 128, 065301 (2022).
\bibitem{lab36a}R. Kubo, Statistical-Mechanical Theory of Irreversible Processes. I. General Theory and Simple Applications to Magnetic and Conduction Problems, J. Phys. Soc. Jpn. 12, 570 (1957).
\bibitem{lab37a}P. C. Martin and J. Schwinger, Theory of Many-Particle Systems. I, Phys. Rev. 115, 1342 (1959).
\bibitem{lab37}Artem Strashko, Peter Kirton, and Jonathan Keeling, Organic Polariton Lasing and the Weak to Strong Coupling Crossover, Phys. Rev. Lett. 121, 193601 (2018).
\bibitem{lab38}Mohammad Ramezani, Quynh Le-Van, Alexei Halpin, and Jaime G\'{e}mez Rivas, Nonlinear Emission of Molecular Ensembles Strongly Coupled to Plasmonic Lattices with Structural Imperfections, Phys. Rev. Lett. 121, 243904 (2018).
\bibitem{lab39}V. Yu. Shishkov, E. S. Andrianov, A. A. Pukhov, A. P. Vinogradov, and A. A. Lisyansky, Zeroth law of thermodynamics for thermalized open quantum systems having constants of motion, Phys. Rev. E 98, 022132 (2018).
\bibitem{lab40}Samuel L. Braunstein, Carlton M. Caves G. J. Milburn, Generalized Uncertainty Relations: Theory, Examples, and Lorentz Invariance, Ann. Phys. (NY) 247, 135 (1996).
\bibitem{lab41}Samuel L. Braunstein and Carlton M. Caves, Statistical distance and the geometry of quantum states, Phys. Rev. Lett. 72, 3439 (1994).
\bibitem{lab42}V. Giovannetti, S. Lloyd, and L. Maccone, Advances in quantum metrology, Nat. Photonics 5, 222 (2011).
\bibitem{lab42a}Vittorio Gorini and Andrzej Kossakowski, Completely positive dynamical semigroups of N-level systems, J. Math. Phys. 17, 821 (1976).
\bibitem{lab43a}H.-P. Breuer, and F. Petruccione, The Theory of Open Quantum Systems (Oxford University Press on Demand,
New York, 2002).
\bibitem{lab43}T. K. Hakala, A. J. Moilanen, A. I. V\"{a}kev¡§ainen, R. Guo, J.-P. Martikainen, K. S. Daskalakis,
H. T. Rekola, A. Julku, and P. T¡§orm¡§a, Bose¨Ceinstein condensation in a plasmonic lattice,
Nature Physics 14, 739 (2018).
\bibitem{lab44}A. I. V\"{a}kev\"{a}inen, A. J. Moilanen, M. Ne\v{c}ada, T. K. Hakala, K. S. Daskalakis, and P. T\"{o}rm\"{a},
Sub-picosecond thermalization dynamics in condensation of strongly coupled lattice plasmons,
Nature communications 11, 1 (2020).



\end{thebibliography}
\end{document}